\documentstyle[epsfig]{elsart}
\begin{document}

\begin{frontmatter}

\title{Excitation of multiple giant dipole resonances:\\
from spherical to deformed nuclei}

\author[Sevilla]{M.V.~Andr\'es},
\author[Sevilla,Catania]{E.G.~Lanza},
\author[GANIL]{P.~Van Isacker},
\author[Orsay]{C.~Volpe}
and
\author[Catania]{F.~Catara}

\address[Sevilla]{Departamento de F\'{\i}sica At\'omica, Molecular y Nuclear,
Universidad de Sevilla, Apdo 1065, 41080 Sevilla, Spain}
\address[Catania]{Dipartimento di Fisica dell'Universit\`a and INFN,
Sezione di Catania, I-95129 Catania, Italy}
\address[GANIL]{GANIL, BP 5027, F-14076 Caen Cedex 5, France}
\address[Orsay]{Groupe de Physique Th\'eorique,
Institut de Physique Nucl\'eaire, F-91406 Orsay Cedex, France}

\begin{abstract}
The effect of deformation on the excitation of multiple giant dipole
resonances is studied.  Analytical expressions are derived in the
framework of the interacting boson model for the energies and E1
properties of giant dipole resonances in spherical and deformed
nuclei, and a numerical treatment of transitional nuclei is proposed.
Coulomb-excitation cross sections are calculated in $^{238}$U and in
the samarium isotopes.
\end{abstract}
\end{frontmatter}

Nuclear multiphonons both at low and at high energies have attracted
much interest lately~\cite{EML94,CHO95,AUM98,WU94}.  At low energy,
the controversy centers around the collective character of multiphonon
states.  While in vibrational nuclei firm experimental evidence now
exists for states with triple quadrupole phonon character (see,
e.g.,~\cite{DEL93}), in deformed nuclei the collective character of
double $\beta$ and $\gamma$ vibrations is still a matter of
acrimonious debate.  At high energy, the study of multiphonons has
pointed out the limitations of the small-amplitude approximation for
vibrational collective motion and of the linear approximation for the
external exciting field.  These assumptions are routinely made for the
first phonon but recent studies of double
phonons~\cite{VOL95,LAN97,LAN98} have shown the importance of
anharmonicities and non-linearities in the excitation of
large-amplitude vibrations.  Similarly, anharmonicities and
non-linearities in nuclear vibrations at low energy have been shown to
play a crucial role in the calculation of the heavy-ion fusion cross
sections at energies close to the Coulomb barrier~\cite{HAG97,HAG98}.

In this letter yet another aspect of phonon excitations at high energy
is investigated, namely the modification of the excitation of the
giant dipole resonance (GDR) away from shell closure as a result of
deformation.  Already at the level of the single GDR the strength
function shows a splitting into two peaks associated with the two
different frequencies of vibration along with or perpendicular to the
axis of axial symmetry.  The question under scrutiny here is how
deformation influences the E1 strength to the double GDR (DGDR) and
what its effect is on the Coulomb excitation cross section.  The
proposed approach makes use of the interacting boson model
(IBM)~\cite{IAC87} for the description of the low-energy collective
levels.  These are coupled to the GDR excitations modelled as $p$
bosons.  The advantage of using the IBM is that the multiphonon states
are obtained as exact eigensolutions of the hamiltonian and,
therefore, no folding procedure is required to obtain the double
phonon states, as is for instance the case in~\cite{PON94,PON98}.  The
folding procedure is only approximately correct for vibrational or
well-deformed nuclei (in the latter case the folding must be done in
the intrinsic frame assuming additivity of intrinsic phonons) but
there is no simple recipe in the intermediate case of transitional
nuclei.  Another advantage of the IBM is that calculations are quick
and this enables an easy estimate of the excitation cross section to
the triple GDR concerning which experiments are currently
planned~\cite{FRApc}.

A general and appropriate basis to discuss the problem of multiple GDR
excitations is of the form
\begin{equation}
|\alpha_LL\times nR;JM_J\rangle.
\label{weak}
\end{equation}
The multiple GDR is built on a low-energy nuclear state characterised
by $\alpha_LL$ where $L$ is the angular momentum of the state and
$\alpha_L$ any other label.  The multiplicity of the GDR is indicated
by $n$ (i.e., $n=1$ for a single GDR, $n=2$ for a double GDR, etc.)
and its angular momentum by $R$.  The single resonance is approximated
as a $p$ boson; the allowed angular momenta of the multiple GDR are
$R=n,n-2,\dots,0$ or 1.  The basis~(\ref{weak}) can be referred to as
a {\it weak-coupling basis} in the sense that the angular momenta $L$
and $R$ are good quantum numbers and are coupled to total angular
momentum $J$ with projection $M_J$.  The weak-coupling basis arises
naturally when the interaction between the GDR and the low-lying
states is weak or is of dipole type $\hat L\cdot\hat R$.  More general
nuclear interactions are not necessarily diagonal in the
basis~(\ref{weak}).  Specifically, the interaction predominantly
responsible for the splitting of the GDR in deformed nuclei is of
quadrupole type $\hat Q_L\cdot\hat Q_R$ and is not diagonal in the
weak-coupling basis~(\ref{weak}).  From the analogous problem in the
particle--core coupling model~\cite{BOH75} it is known that the
diagonalisation of $\hat Q_L\cdot\hat Q_R$ in the basis~(\ref{weak})
gives rise to a {\it strong-coupling basis} with the quantum
numbers~\cite{BRI87}
\begin{equation}
|\alpha_L\times n;\alpha_JK_JJM_J\rangle.
\label{strong}
\end{equation}
The angular momenta $L$ and $R$ no longer are conserved quantities and
are replaced by $K_J$, the projection of the total angular momentum
$J$ on the axis of axial symmetry.  The choice of basis depends on the
competition between various terms in the nuclear hamiltonian.  Large
splittings in $L$ or $R$ induce the weak-coupling basis~(\ref{weak});
if, in contrast, these are small in comparison with the quadrupole
coupling between the low-energy states and the GDR, such as is the
case in well-deformed nuclei, the strong-coupling basis~(\ref{strong})
is obtained.

The above remarks are rather general and model independent.
Low-energy nuclear states and their coupling with the GDR can be
modelled in several different ways and a convenient one is in the
context of the IBM~\cite{IAC87,IAC91}.  In deformed nuclei the problem
was worked out analytically for a single GDR by Rowe and
Iachello~\cite{ROW83}, while numerical results were presented for the
single GDR in~\cite{MOR82,SCH83,MAI84,ZUF87} and later for the double
GDR in~\cite{SCH90}.  In this letter analytical results for the
energies and E1 transitions in spherical and deformed nuclei are
generalised to the multiple GDR, and numerical results are presented
for intermediate cases. From these results the Coulomb-excitation
cross sections are calculated taking into account the dynamics of
the collision, which was lacking in previous IBM treatments of the
single and the double GDR.

It is assumed in the IBM that collective nuclear states can be
described in terms of $N$ $s$ and $d$ bosons where $N$ is half the
number of valence nucleons~\cite{IAC87}.  The dynamical algebra of the
model is U(6) in the sense that a single of its representations
(namely the symmetric one, $[N]$) is assumed to contain all low-energy
collective nuclear states.  To this space are coupled the multiple GDR
excitations.  Assuming that a single GDR is described by a $p$ boson,
multiple GDR excitations are represented by the direct sum
$(0,0)\oplus(1,0)\oplus(2,0)\oplus\cdots$ of symmetric representations
$(n,0)$ of U(3).  The dynamical algebra of the coupled system is thus
${\rm U}(6)\otimes{\rm U}(3)$ with the proviso that several U(3)
representations must be taken to build the model space.

The model hamiltonian has the generic form~\cite{IAC91}
\begin{equation}
\hat H=
\hat H_{sd}+
\hat H_p+
\hat V_{sd-p}.
\label{ham}
\end{equation}
The first term is the usual IBM hamiltonian~\cite{IAC87} which gives
an adequate description of the low-energy spectrum of spherical,
deformed and transitional nuclei.  The second term in~(\ref{ham})
governs the multiple GDR spectrum and is of the form
\begin{equation}
\hat H_p=
\epsilon_p\hat n+
\alpha_p\hat n(\hat n+3)+
\beta_p\hat R^2,
\label{ham_p}
\end{equation}
where $\hat n$ is the $p$-boson number operator and $\hat R$ the
associated angular momentum operator.  The coefficient $\epsilon_p$
represents the unperturbed single GDR energy.  The second term
in~(\ref{ham_p}) induces a diagonal anharmonicity in the excitation
energy of the multiple GDR.  The interaction term in~(\ref{ham}),
finally, acquires the form
\begin{equation}
\hat V_{sd-p}=
\alpha_0\hat n_d\hat n+
2\alpha_1\hat L\cdot\hat R+
2\sqrt{3}\,\alpha_2\hat Q_L\cdot\hat Q_R,
\label{ham_sdp}
\end{equation}
with a monopole, dipole and quadrupole interaction.  More complicated,
higher-order interactions that mix excitations with different $n$ are
not included here.  The hamiltonian~(\ref{ham}) can be diagonalised
numerically in the ${\rm U}(6)\otimes{\rm U}(3)$ model space.  For
particular values of the parameters analytical solutions are
available.  Two classes of analytically solvable hamiltonians exist.

\noindent
{\it (i) Weak-coupling limit.}  A weak-coupling hamiltonian can be
defined for each of the three limits [U(5), SU(3) and O(6)] of the IBM
and is obtained for $\alpha_0=\alpha_2=0$ in~(\ref{ham_sdp}).  In the
particular case of the U(5) limit, only $\alpha_2=0$ is required.  In
the U(5) weak-coupling limit states are labelled by
\begin{equation}
|[N]n_d vn_\Delta L\times(n,0)R;JM_J\rangle,
\label{weaklab}
\end{equation}
where $n_d$ is the number of $d$ bosons, $v$ the $d$-boson seniority
and $n_\Delta$ an additional quantum number related to the pairing of
triplets of $d$ bosons \cite{IAC87,CAS79}.  The energy eigenvalues of
the states~(\ref{weaklab}) are
\begin{eqnarray}
E&=&
\epsilon_d n_d+
\alpha_d n_d(n_d+4)+
\gamma_d v(v+3)+
(\beta_d -\alpha_1) L(L+1)+
\nonumber\\&&
\epsilon_p n+
\alpha_p n(n+3)+
(\beta_p -\alpha_1) R(R+1)+
\alpha_0 n_d n +
\alpha_1 J(J+1),
\label{weakeig}
\end{eqnarray}
where $\beta_d$ is the parameter associated with $\hat L^2$ in $\hat
H_{sd}$.

\noindent
{\it (ii) Strong-coupling rotational limit.}  The appropriate
classification of the low-energy states in this case is SU(3).  This
symmetry requires that the quadrupole operator $\hat Q_L$ be a
generator of SU(3) and that $\alpha_0=0$ and
$\beta_p=\beta_d=\alpha_1-{3\over4}\alpha_2$, In the SU(3)
strong-coupling limit states are labelled by
\begin{equation}
|[N](\lambda_{sd},\mu_{sd})\times(n,0);
(\lambda,\mu)K_JJM_J\rangle.
\label{stronglab}
\end{equation}
The labels $(\lambda_{sd},\mu_{sd})$ are associated with the SU(3)
algebra of the $s$ and $d$ bosons.  They characterise the band
structure of the low-energy spectrum; for example, the ground band has
$(\lambda_{sd},\mu_{sd})=(2N,0)$.  The energy eigenvalues of the
states~(\ref{stronglab}) are
\begin{eqnarray}
E&=&
(\alpha_{sd}-\alpha_2)
[\lambda_{sd}(\lambda_{sd}+3)
+\mu_{sd}(\mu_{sd}+3)
+\lambda_{sd}\mu_{sd}]+
\nonumber\\&&
\epsilon_p n+
(\alpha_p-\alpha_2)n(n+3)+
\nonumber\\&&
\alpha_2[\lambda(\lambda+3)+\mu(\mu+3)+\lambda\mu]+
(\alpha_1-{\textstyle{3\over4}}\alpha_2)J(J+1).
\label{strongeig}
\end{eqnarray}
The splitting of the GDR comes about because of the fourth term
in~(\ref{strongeig}).  The $({\rm GDR})^n$ excitation splits into
$n+1$ peaks corresponding to $(\lambda,\mu)$ values
\begin{equation}
(\lambda,\mu)=(2N+n,0),(2N+n-2,1),\dots,(2N-n,n),
\label{muln}
\end{equation}
where $2N\geq n$ is assumed.  The energies of the different peaks are
found from~(\ref{strongeig}). The values of $K_J$ and $J$ allowed for
a given $(\lambda,\mu)$ representation are given by Elliott's
rule~\cite{IAC87,ELL58}.

The Coulomb excitation of GDRs occurs predominantly through E1.  In
the context of the present model an E1 excitation corresponds to the
creation of a $p$ boson (annihilation in case of E1 de-excitation) and
thus the electric multipole operator ${\cal M} ({\rm E}1\mu)$~\cite{BOH75}
is parametrised as $\zeta(p^\dag_\mu+\tilde p_\mu)$.  The calculation
of E1 transition probabilities requires the matrix elements of
$p^\dagger$ in the basis~(\ref{weaklab}) or~(\ref{stronglab}) which
can be done by standard group-theoretical techniques.  Analytical
expressions are found in the two limiting cases.

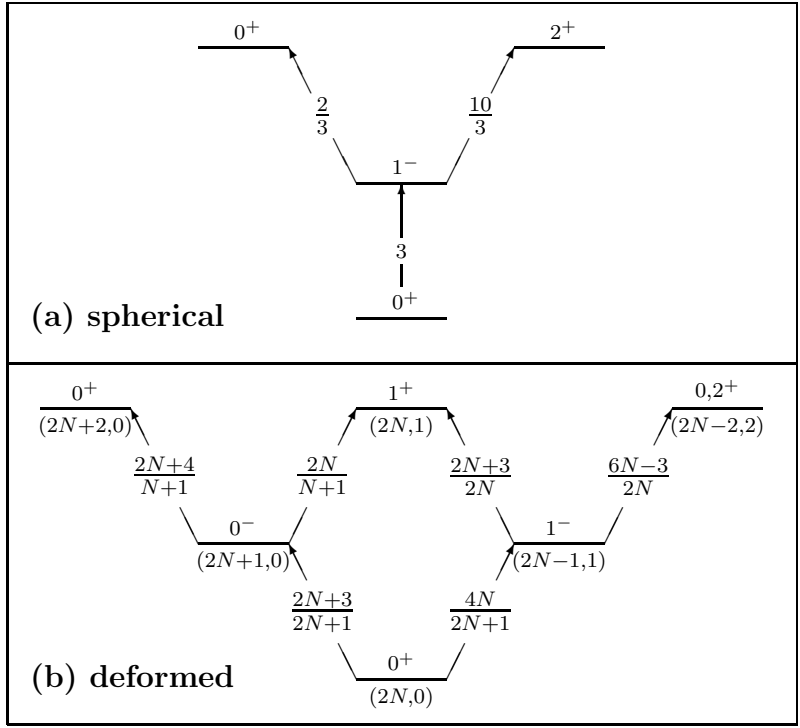
\begin{figure}
\centering
\setlength{\unitlength}{0.6mm}
\begin{picture}(175,160)(-5,0)
\put( -5,  0){\line(1,0){175}}
\put( -5, 80){\line(1,0){175}}
\put( -5,160){\line(1,0){175}}
\put( -5,  0){\line(0,1){160}}
\put(170,  0){\line(0,1){160}}
\put(  0, 90){\makebox(0,0)[l]{\bf (a) spherical}}
\put(  0, 10){\makebox(0,0)[l]{\bf (b) deformed}}

\put(30,80){
\begin{picture}(100,80)(-5,-10)
\thicklines

\put( 35, 0){\line(1,0){20}}
\put( 45, 4){\makebox(0,0){$\;\scriptstyle{0^+}$}}
\put( 35,30){\line(1,0){20}}
\put( 45,34){\makebox(0,0){$\;\scriptstyle{1^-}$}}
\put(  0,60){\line(1,0){20}}
\put( 10,64){\makebox(0,0){$\;\scriptstyle{0^+}$}}
\put( 70,60){\line(1,0){20}}
\put( 80,64){\makebox(0,0){$\;\scriptstyle{2^+}$}}

\thinlines
\put(45, 7){\line(0,1){5}}
\put(45,15){\makebox(0,0){$\scriptstyle{3}$}}
\put(45,18){\vector(0,1){12}}
\put(35,30){\line(-1,2){5}}
\put(27.5,45){\makebox(0,0){${2\over3}$}}
\put(25,50){\vector(-1,2){5}}
\put(55,30){\line(1,2){5}}
\put(62.5,45){\makebox(0,0){${{10}\over3}$}}
\put(65,50){\vector(1,2){5}}
\end{picture}}

\put(0,0){
\begin{picture}(160,80)(0,-10)
\thicklines
\put( 70, 0){\line(1,0){20}}
\put( 80,-4){\makebox(0,0){$\scriptstyle{(2N,0)}$}}
\put( 80, 4){\makebox(0,0){$\scriptstyle{0^+}$}}
\put( 35,30){\line(1,0){20}}
\put( 45,26){\makebox(0,0){$\scriptstyle{(2N+1,0)}$}}
\put( 45,34){\makebox(0,0){$\scriptstyle{0^-}$}}
\put(105,30){\line(1,0){20}}
\put(115,26){\makebox(0,0){$\scriptstyle{(2N-1,1)}$}}
\put(115,34){\makebox(0,0){$\scriptstyle{1^-}$}}
\put(  0,60){\line(1,0){20}}
\put( 10,56){\makebox(0,0){$\scriptstyle{(2N+2,0)}$}}
\put( 10,64){\makebox(0,0){$\scriptstyle{0^+}$}}
\put( 70,60){\line(1,0){20}}
\put( 80,56){\makebox(0,0){$\scriptstyle{(2N,1)}$}}
\put( 80,64){\makebox(0,0){$\scriptstyle{1^+}$}}
\put(140,60){\line(1,0){20}}
\put(150,56){\makebox(0,0){$\scriptstyle{(2N-2,2)}$}}
\put(150,64){\makebox(0,0){$\scriptstyle{0,2^+}$}}

\thinlines
\put(70,0){\line(-1,2){4}}
\put(62.5,15){\makebox(0,0){${{2N+3}\over{2N+1}}$}}
\put(59,22){\vector(-1,2){4}}

\put(90,0){\line(1,2){4}}
\put(97.5,15){\makebox(0,0){${{4N}\over{2N+1}}$}}
\put(101,22){\vector(1,2){4}}

\put(35,30){\line(-1,2){4}}
\put(27.5,45){\makebox(0,0){${{2N+4}\over{N+1}}$}}
\put(24,52){\vector(-1,2){4}}

\put(55,30){\line(1,2){4}}
\put(62.5,45){\makebox(0,0){${{2N}\over{N+1}}$}}
\put(66,52){\vector(1,2){4}}

\put(105,30){\line(-1,2){4}}
\put(97.5,45){\makebox(0,0){${{2N+3}\over{2N}}$}}
\put(94,52){\vector(-1,2){4}}

\put(125,30){\line(1,2){4}}
\put(132.5,45){\makebox(0,0){${{6N-3}\over{2N}}$}}
\put(136,52){\vector(1,2){4}}
\end{picture}}
\end{picture}
\caption{ E1 excitation patterns up to $({\rm GDR})^2$ in (a) the
spherical weak-coupling limit and (b) the deformed strong-coupling
limit.  (a) The levels are labelled by $J^\pi$ on top.  The numbers
between levels are the strengths $B({\rm
E}1;(n,0)J\rightarrow(n+1,0)J')$ in units of $\zeta^2$.  (b) The
levels are labelled by $(\lambda,\mu)$ underneath and by $K^\pi$ on
top.  The expressions between levels are the summed strengths
$\sum_{K'L'}B({\rm E}1;
(\lambda,\mu)KL\rightarrow(\lambda',\mu')K'L')$ in units of
$\zeta^2$.}
\label{e1}
\end{figure}

\noindent
{\it (i) Weak-coupling limit.}  For the GDR excitations built on the
$0^+$ ground state results up to the DGDR are shown in Fig.~\ref{e1}a.
Generally, for the $B$(E1) values between multipole GDR excitations
built on the $0^+$ ground state one recovers the independent-quanta
result~\cite{BOH75}
\begin{eqnarray}
\lefteqn{
\sum_f
B({\rm E}1;n_d=0\times(n,0)R_i=J_i;J_i\to
           n_d=0\times(n-1,0)R_f=J_f;J_f)}
\nonumber\\&=&
\sum_f
B({\rm E}1;\alpha nJ_i\to\alpha(n-1)J_f)=
nB({\rm E}1;\alpha n=1\to\alpha n=0),
\end{eqnarray}
where $\alpha$ denotes all other quantum numbers that cannot change.

\noindent
{\it (ii) Strong-coupling rotational limit.}  Results up to the DGDR
are shown in Fig.~\ref{e1}b.  Although all individual $B({\rm E}1)$s
are known, for simplicity of presentation only the summed strengths
$\sum_{K'L'}B({\rm E}1;
(\lambda,\mu)KL\rightarrow(\lambda',\mu')K'L')$ are shown.  In the
limit of large boson number $N$ one recovers harmonic results that
have a simple geometric interpretation.  For example, the $B({\rm
E}1)$ values from the $0^+$ ground state to the two $1^-$ GDRs are 1
and 2, respectively, the first associated with an oscillation along
the axis of symmetry (say the $z$ direction) and the second with
oscillations in the $x$ and $y$ directions.  For the single-to-double
GDR excitation one finds $B({\rm E}1)$ values which are, for
$N\rightarrow\infty$, 2, 2, 1 and 3.  The large-$N$ results can be
generalised to (GDR)$^n$.

Coulomb excitation in heavy-ion collisions is usually described by
treating the relative motion classically while the internal degrees of
freedom of the colliding nuclei are accounted for quantum
mechanically.  The operator responsible for the excitation depends on
time through the relative distance.  For relativistic collisions its
expression is as in equation~(35) of~\cite{LAN97} where each term of
the multiple expansion of the external field factorizes into two
elements. The first depends on the collision properties, the second on
the structure of the nucleus being excited. In the present study only
contributions from the ${\cal M}({\rm E}1)$ matrix elements are
considered and those are calculated within the model described above.
The solution of the time-dependent Schr\"odinger equation leads to a
set of coupled differential equations for the probability amplitudes
to excite the (GDR)$^n$ states.  For each impact parameter $b$ these
equations are integrated along the appropriate classical trajectory.
For each (GDR)$^n$ state the total inelastic cross section is then
obtained by integrating the corresponding probability over all impact
parameters, starting from a mininum value $b_{\rm min}=1.34
[A_1^{1/3}+A_2^{1/3} -0.75 (A_1^{-1/3}+A_2^{-1/3})]$~fm~\cite{BEN89}.
 
The above formalism will now be applied to the relativistic Coulomb
excitation of the single and double GDRs in $^{238}$U and in the chain
of even isotopes $^{148}$Sm to $^{154}$Sm.  The former is an example
of a well-deformed nucleus while the samarium isotopes exhibit a
change from vibrational to deformed as $A$ increases.  The description
of such structural changes requires the use of a transitional IBM
hamiltonian and hence the following analysis is not confined to any of
the previously discussed analytical limits but always involves a
numerical diagonalisation.  For $^{238}$U, a
consistent-$Q$~\cite{WAR83} hamiltonian $\hat H_{sd}= \kappa\hat
Q^\chi\cdot\hat Q^\chi+ \kappa'\hat L\cdot\hat L$ is used with
$\kappa=-16$~keV, $\kappa'=1.5$~keV and $\chi=-0.72$.  These
parameters yield an adequate description of the ground--gamma band
splitting, of the moments of inertia and of the E2 transitions from
gamma to ground band.  The $\hat H_{sd}$ hamiltonian for the Sm
isotopes is taken as in~\cite{MAI84}.  The additional parameters
$\epsilon_p$, $\alpha_0$, $\alpha_2$ and $\zeta$ in $\hat H_p$, $\hat
V_{sd-p}$ and in the E1 operator are given in Table~\ref{param}.  They
have been chosen as to reproduce the observed photoabsorption cross
section~\cite{phou,phosm} to the first GDR.  Agreement is obtained if
to each eigenstate is associated a spreading width $\Gamma_i=0.007
E_i^{2.5}$ in $^{238}$U and $\Gamma_i=0.029 E_i^{1.81}$ in the Sm
isotopes (with $E_i$ and $\Gamma_i$ in MeV). To reproduce the
photoabsorption cross section of a well-deformed nucleus one uses
the fact that the energy splitting of the GDR is very sensitive
to the parameter $\alpha_2$. Then, once $\alpha_2$ is fixed, one varies
$\alpha_0$ to sligthly change the contribution of each peak to the
energy-weigthed sum rule and $\epsilon_p$ to shift the energy of the
dipole states. Finally, the parameter $\zeta$ is obtained from
a global normalisation.

\begin{figure}
\begin{center}
\mbox{\epsfig{file=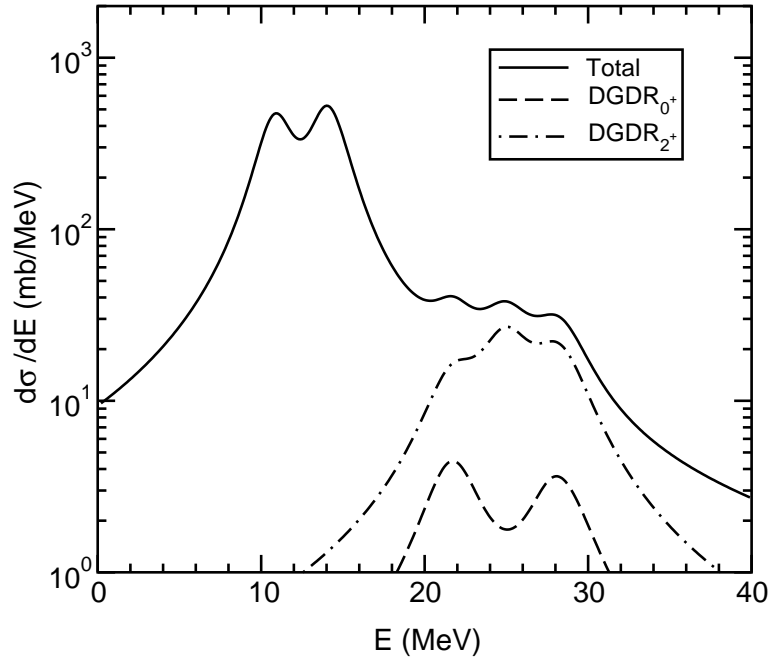,height=15.0truecm,angle=-90}}
\caption{Coulomb excitation of the single and double GDR in $^{238}$U
in the $^{238}$U (0.5 GeV/A) + $^{208}$Pb reaction.
The dashed line is the contribution from the $0^+$ component of the
two-phonon state while the $2^+$ one is represented by the dot--dashed
line.  The results corresponding to the $1^+$ component are too small
to be seen.}
\label{figupb}
\end{center}
\end{figure}

Recently, an experiment was done for $^{238}$U + $^{208}$Pb  at
0.5~GeV/A (the data analysis is in progress~\cite{expu}). The result
of the corresponding calculation is shown in Fig.~\ref{figupb}.  The
full line corresponds to the total cross section obtained by smoothing
the cross section to each discrete state with a Lorentzian having a
width of $\Gamma_1=2.5$~MeV and $\Gamma_2=\sqrt2\Gamma_1$ for the
single and double GDR states, respectively. As expected, the GDR peak
is split in two while a broad plateau occurs in the DGDR region.  The
integrated cross sections are $\sigma^{\rm GDR}=$ 3.5 b and
$\sigma^{\rm DGDR}=$ 0.3 b, lower than those of Ponomarev
{\it et al.}~\cite{PON98}. The difference can be ascribed to the fact that
a coupled-channel method is used while in ref.~\cite{PON98} the cross
sections are calculated in first- and second-order perturbation theory. In
fact, if we use perturbation theory then the result for the
$\sigma^{\rm GDR}$ increases up to 4.2 b. The two approaches give
similar results only for large impact parameters. Some difference
comes also from the different $B$(E1) distribution.
Figure~\ref{figupb} also shows the contributions associated with the $0^+$
(dashed line) and $2^+$ (dot-dashed line) component of the DGDR.  The
$1^+$ component does not appear in the figure since it is extremely
small.  After substraction of the long single GDR tail, the three
peaks, expected from~(\ref{muln}), are clearly visible.  (If the
convolution of the cross section were done with a
$\Gamma_2=2\Gamma_1$, the three peaks are still visible though less
evident).  Therefore, an exclusive experiment in coincidence with
$\gamma$--$\gamma$ decay might conceivably give a direct signature of
the excitation of the DGDR.

The present results differ from those of Ponomarev {\it et
al.}~\cite{PON98} who study the same reaction with the
particle--phonon model, using second-order perturbation theory to
calculate the cross section.  In ref.~\cite{PON98} the DGDR cross
section appears as a structureless peak while here it does not.  The
reason is that the calculated cross section to the single GDR in
ref.~\cite{PON98} shows three peaks in the energy region 11--15~MeV.
As a result, since the DGDR states are constructed as products of two
single GDR states, one expects six peaks in the DGDR energy region
which eventually smear out the cross section to the DGDR. In our case
we have only two peaks, which are present in the experimental data,
because we have fixed the parameters of our hamiltonian by fitting the
photoabsorption cross section.

\begin{figure}
\begin{center}
\mbox{\epsfig{file=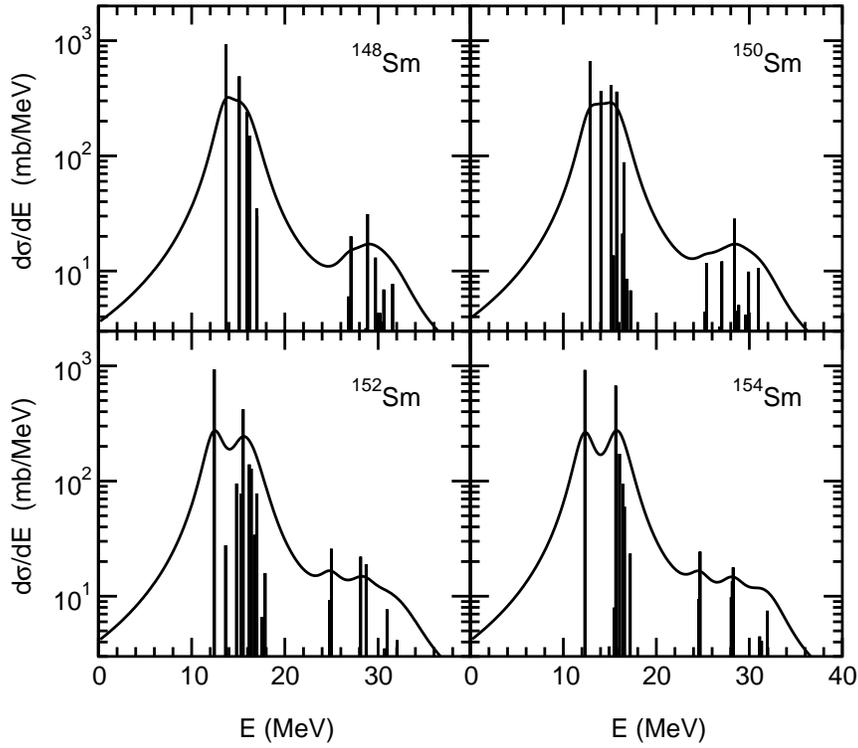,height=16.0truecm,angle=-90}}
\caption{Coulomb excitation cross sections of single and double GDR in
several samarium isotopes in the $^{208}$Pb (0.5 GeV/A) + $^A$Sm
reaction. The bars correspond to the cross sections to the discrete
states and the full line corresponds to the convolution of these cross
sections by a lorentzian of width $\Gamma_1=2.5$~MeV for the
one-phonon states and $\Gamma_2=\sqrt{2}\Gamma_1$ for the two-phonon
states.}
\label{figsm}
\end{center}
\end{figure}

The results of the calculated inelastic cross sections for the
reactions $^{208}$Pb + $^A$Sm at 0.5~GeV/$A$ are shown in
Fig.~\ref{figsm}.  The cross sections to each discrete state are shown
as well as the ones obtained by the same smoothing procedure described
above.  In the transition from spherical to deformed one observes a
continous evolution in the shape of the cross section resulting in the
deformed case in a clear splitting in two peaks of the GDR and,
correspondingly, three bumps in the DGDR energy region.  This effect
is due to the increase in separation between the two main components
of the single GDR and the concentration of the small components in a
more narrow energy range which in turn results from the increasing
coupling of the GDR with quadrupole modes.

In summary, energy and E1 properties of multiple giant dipole
resonances have been studied in the context of the interacting boson
model.  The model has been applied to $^{238}$U and to transitional
samarium isotopes for which Coulomb excitation cross sections have
been calculated in the reaction with $^{208}$Pb at 0.5~GeV/$A$.  The
example of the samarium isotopes shows how the excitation cross
section is modified when going from spherical to deformed nuclei.
The calculation shows also that exclusive experiments on a well-deformed
nucleus like $^{238}$U could give a direct signature of the existence
of the double giant dipole resonance.

This work has been supported by the Spanish DGICyT under contract
PB95-0533-A, by an agreement between IN2P3 (France) and CICYT (Spain)
and by an agreement between INFN (Italy) and CICYT (Spain). E.G.L. is a
Marie Curie Fellow with contract ERBFMBICT983090 within the TMR
program of the European Community.


\newpage

\begin{table}
\caption{Parameters of the hamiltonian
and of the E1 transition operator.}
\label{param}
\begin{tabular}{rcrcrcrcr}
\hline\hline
Isotope&&$\epsilon_p$~(MeV)&&$\alpha_0$~(keV)&&
$2\sqrt{3}\,\alpha_2$~(keV)&&$\zeta$~(e fm)\\
\hline
$^{148}$Sm &&14.84&&  0&&$-275$&&3.84\\
$^{150}$Sm &&14.59&&  0&&$-275$&&3.84\\
$^{152}$Sm &&14.05&&250&&$-275$&&3.84\\
$^{154}$Sm &&13.99&&150&&$-225$&&3.84\\
$^{238}$U  &&13.00&&  0&&$-160$&&5.06\\
\hline\hline
\end{tabular}
\end{table}


\begin{thebibliography}{99}

\bibitem{EML94}
H.~Emling, 
Prog.\ Part.\ Nucl.\ Phys.\ {\bf33} (1994) 729.

\bibitem{CHO95} 
Ph.~Chomaz and N.~Frascaria, 
Phys.\ Rep.\ {\bf 252} (1995) 275.

\bibitem{AUM98} 
T.~Aumann, P.F.~Bortignon and H.~Emling,
Annu.\ Rev.\ Nucl.\ Part.\ Sci.\ {\bf48} (1998) 351.

\bibitem{WU94}
X.~Wu, A.~Aprahamian, S.M.~Fischer, W.~Reviol, G.~Liu
and J.X.~Saladin,
Phys.\ Rev.\ C {\bf49} (1994) 1837.

\bibitem{DEL93}
M.~D\'el\`eze, S.~Drissi, J.~Jolie, J.~Kern and J.P.~Vorlet,
Nucl.\ Phys.\ A {\bf554} (1993) 1.

\bibitem{VOL95}
C.~Volpe, F.~Catara, Ph.~Chomaz, M.V.~Andr\'es and E.G.~Lanza,
Nucl.\ Phys.\ A {\bf589} (1995) 521.
                 
\bibitem{LAN97}
E.G.~Lanza, M.V.~Andr\'es, F.~Catara, Ph.~Chomaz and C.~Volpe,
Nucl.\ Phys.\ A {\bf613} (1997) 445.

\bibitem{LAN98}
E.G.~Lanza, M.V.~Andr\'es, F.~Catara, Ph.~Chomaz and C.~Volpe, 
Nucl.\ Phys.\ A {\bf636} (1998) 452.
               
\bibitem{HAG97}
K.~Hagino, N.~Takigawa, M.~Dasgupta, D.J.~Hinde and J.R.~Leigh,
J.\ Phys.\ G {\bf23} (1997) 1413.

\bibitem{HAG98}
K.~Hagino, S.~Kuyucak and N.~Takigawa,
Phys.\ Rev.\ C {\bf57} (1998) 1349.

\bibitem{IAC87}
F.~Iachello and A.~Arima,
{\it The Interacting Boson Model}
(Cambridge University Press, Cambridge, 1987).

\bibitem{PON94}
V.Yu.~Ponomarev et al., 
Phys.\ Rev.\ Lett.\ {\bf72} (1994) 1168.

\bibitem{PON98}
V.Yu.~Ponomarev, C.A.~Bertulani and A.V.~Sushkov,
Phys.\ Rev.\ C {\bf58} (1998) 2750.

\bibitem{FRApc}
N.~Frascaria, private communication.

\bibitem{BOH75}
A.~Bohr and B.R.~Mottelson,
{\it Nuclear Structure. I and II.}
(Benjamin, Reading, 1975).

\bibitem{BRI87}
D.M.~Brink, B.~Buck, R.~Huby, M.A.~Nagarajan and N.~Rowley,
J.\ Phys.\ G {\bf13} (1987) 629.

\bibitem{IAC91}
F.~Iachello and P.~Van Isacker,
{\it The Interacting Boson--Fermion Model}
(Cambridge University Press, Cambridge, 1991).

\bibitem{ROW83}
D.J.~Rowe and F.~Iachello,
Phys.\ Lett.\ B {\bf130} (1983) 231.

\bibitem{MOR82}
I.~Morrison and J.~Weise,
J.\ Phys.\ G {\bf8} (1982) 687.

\bibitem{SCH83}
F.G.~Scholtz and F.J.W.~Hahne,
Phys.\ Lett.\ B {\bf123} (1983) 147.

\bibitem{MAI84}
G.~Maino, A.~Ventura, L.~Zuffi and F.~Iachello,
Phys.\ Rev.\ C {\bf30} (1984) 2101.

\bibitem{ZUF87}
L.~Zuffi, P.~Van Isacker, G.~Maino and A.~Ventura,
Nucl.\ Instr.\ Meth.\ Phys.\ Res.\ A {\bf255} (1987) 46.

\bibitem{SCH90}
F.G.~Scholtz and F.J.W.~Hahne,
Z.\ Phys.\ A {\bf336} (1990) 145.

\bibitem{ELL58}
J.P.~Elliott,
Proc.\ R.\ Soc.\ London A {\bf245} (1958) 562.

\bibitem{CAS79}
O.~Casta\~nos, E.~Chac\'on, A.~Frank and M.~Moshinsky,
J.\ Math.\ Phys.\ {\bf20} (1979) 35.

\bibitem{BEN89}
C.J.~Benesh, B.C.~Cook and J.P.~Vary,
Phys.\ Rev.\ C {\bf40} (1989) 1198.

\bibitem{WAR83}
D.D.~Warner and R.F.~Casten,
Phys.\ Rev.\ C {\bf28} (1983) 1798.

\bibitem{phou}
G.M.~Gurevich et al, Nucl.\ Phys.\ A {\bf273} (1976) 326.\\
A.~Veyssiere et al, Nucl.\ Phys.\ A {\bf119} (1973) 45.\\
J.T.~Caldwell et al, Phys.\ Rev.\ C {\bf21} (1980) 1215.


\bibitem{phosm}
P.~Carlos et al, Nucl.\ Phys.\ A {\bf225} (1974) 171.


\bibitem{expu}
H.~Emling, private communication.

\end{thebibliography}
\end{document}